\documentstyle[12pt]{article}
\begin{document}
\title{Integrable hierarchies 
from BRST-anti-BRST gauge-fixing}
\author{V. Calian
\\Niels Bohr Institute, Blegdamsvej 17 
\\DK-2100 Copenhagen, Denmark
\\and University of Craiova, Faculty of Physics, 13 A. I. Cuza 
\\Craiova 1100, Romania}
\date{}
\maketitle
\begin{abstract}
The BRST formulation is used in order to derive the existence criterion for classical 
bi-Hamiltonian systems, based on non-anomalous deformation of the gauge-fixing structure. The 
recursion operator 
is then used to provide the entire hierarchy of integrable models associated to the original 
BRST and anti-BRST charges.
\end{abstract}


\section{Introduction}
An important class of theories, including topological field theories (as in \cite{13}), 
reparametrization invariant models \cite{7}, even the (pseudo-) classical mechanics \cite{7}-\cite{9} in
BRST-anti-BRST formulation are confronted to a common problem: the
so-called zero-Hamiltonian, which is also hoped to establish the 
corresponding connections between them.
In all of these cases, inequivalent theories are generated when using 
different gauge-fixing ``fermions'' which can not be continuously deformed one 
into the other. 

In the BRST Lagrangian action, the gauge-fixing term can be written 
as $s_1 \psi$ or, when the anti-BRST symmetry is taken into account, 
as $s_1s_2B$. However, if one starts with the BRST-anti-BRST path-integral \cite{1}-\cite{5}
 of the (-pseudo)classical mechanics, the function $B$ is exactly the 
original Hamiltonian $H$ of the system and it entirely determines its 
time-evolution, i.e. the equations of motion 
$\dot{\phi}^a=\omega^{ab} \partial_b H$ (with $a=1,2n$).

In this paper, we will find the answer to a new question, which is also common to the set of 
interrelated models mentioned above: is the BRST framework able to give a
criterion to separate the integrable and non-integrable models? 
For the beginning, we will address the Liouville complete integrability and we will obtain the condition for a bi-Hamiltonian structure to exist. The recursion operator, the compatibility of
the Poisson brackets and the entire integrable hierarchy are then constructed, using 
the Hamiltonian BRST-anti-BRST formalism.  
Moreover, the quantization stage for the integrable models becomes systematic and eliminates the well known difficulties of the very a few documented attempts \cite{11}.

In section 2, the bi-Hamiltonian structure is derived as a consequence of the simultaneous deformations of the anti-BRST charge $\overline{Q}$  and ``gauge-fixing 
function''
 $H( \phi )$ which leave the effective Hamiltonian invariant. As a result, the recursion operator, the conserved functionals, the associated hierarchies and an explicit example are given in the section 3. The paper concludes (section 4) with a 
cohomological interpretation, a discussion of the quantization stage and possible 
extensions of the method to a ``multi-level'' Hamiltonian formalism or to the constrained classical Hamiltonians.

\section{Gauge-fixing  and bi-Hamiltonian structures}
The usual invariance of the path integral at gauge-fixing changes is not recovered in the BRST formulation of the classical mechanics, where an abelian gauge theory is obtained, the phase space coordinates $\phi$ being left completely arbitrary, and the Hamilton's equation are obtained by the choice of $\psi$.

However, since $\psi=\{ H, \overline{Q} \}$, a complementary problem has to be solved: is it possible to find simultaneous deformations of both $H$ and $\overline{Q}$, such that the same effective Hamiltonian $H_{eff} =- \{ Q, \{ \overline{Q}, H \}\}$ and the same symmetry algebra are generated. 

If the deformation exists ($\{Q,\psi_1-\psi\}=0$), we will show that a bi-Hamiltonian structure is obtained. On the opposite, if the theory is ``anomalous'', in this extended sense, the system is not integrable.

The general results of \cite{7}-\cite{9} are the starting point of our analysis, but the procedure proposed here may be applied to any theory where the effective Hamiltonian is written as a BRST-anti-BRST invariant or at least cancels on the surface defined by the equations of motion. 

It is useful to remark that, due to the structure of the extended algebra $Q, \overline{Q}, K, 
\overline{K}, Q_g$, to the explicit expression of the gauge-fixing term, and to the condition 
that $\psi_1-\psi$ must be BRST-exact, the deformation of the anti-BRST charge is given, in 
a first order, by: 
\begin{equation}
\overline{Q}_1=\overline{Q}+\delta \overline{Q}
\label{1}
\end{equation}
where the BRST-exact deformation may be expressed as:
\begin{equation}
\delta\overline{Q}=\{ \overline{K}_1,Q \}
\label{2}
\end{equation}
with the definition:
\begin{equation}
\overline{K}_1={\cal{P}}_a \omega_1^{ab} {\cal{P}}_b 
\label{3}
\end{equation}

In order to obtain the same $H_{eff}$ for the ``dressed'' bicomplex,to preserve its character 
and to keep the conservation of the new BRST and anti-BRST charges, the following 
conditions should be satisfied:
\begin{equation}
\{\overline{Q}_1,\overline{Q}_1\}=0; \{Q_1,Q_1\}=0; \{\overline{Q}_1, Q_1\}=0 
\label{4}
\end{equation}
\begin{equation}
\{\overline{Q}_1,H_{eff}\}=\{Q_1,H_{eff}\}=0; \{\overline{Q}_1,H_1\}=\{\overline{Q},H\}
\label{5}
\end{equation}

The computation of the new $Q_1$ as $\{K_1,\overline{Q}_1\}$ leads to the expected $Q_1=Q$, 
while the explicit form of $\delta \overline{Q}$ in (\ref{2}) becomes:
\begin{equation}
\delta\overline{Q}=-{\cal{P}}_b \lambda_a \left(\omega_1-\omega\right)^{ab}
\label{6}
\end{equation}
A straightforward calculation of the conditions (\ref{4}), (\ref{5}) enforces that:
\begin{equation}
\omega_1^{ab}\partial_b H = \omega^{ab}\partial_b H_1
\label{7}
\end{equation} 
and the compatibility of the two Poisson brackets:
\begin{equation}
\sum\limits_{\{abc\}}\left(\omega^{am}\partial_m \omega_1^{bc}+
\omega_1^{am}\partial_m \omega^{bc}\right)=0
\label{8}
\end{equation}
which are exactly the conditions for a bi-Hamiltonian structure to exist:
\begin{equation}
J \delta H_1=J_1\delta H
\label{9}
\end{equation}

It is also a matter of direct calculation to verify that the enlarged symmetry algebra defined by $Q, \overline{Q}_1, K, K_1, \overline{K}_1, Q_g$ is not broken. 

Several consequences of this analysis should be mentioned: 

i)if one acts only on the original Hamiltonian and leaves the Poisson structure, and thus the anti-BRST charge, undeformed, our previous conditions degenerate to imposing that $H_1\left(\phi\right)$ are anti-BRST invariants and the action is modified by a trivial term. On the other hand, the enlarged symmetry is preserved only if the BRST bracket of $H$ and $H_1$ is zero, and $H_1$ is a trivial integral of motion.

ii)if we perform only a deformation of the symplectic tensor and keep the $H$ unmodified, our condition becomes:
\begin{equation}
{\cal{P}}_a \omega_1^{ab}\partial_b H=0
\label{10}
\end{equation} 
which may be interpreted as:
\begin{equation}
\{ \overline{Q}_1, H \}=0
\label{11}
\end{equation}
such that $H$ is an anti-BRST invariant of a different theory, defined by the symplectic 
tensor $\omega_1$. Again, the (non) conservation of the enlarged algebra would 
lead us to compatibility of the Poisson brackets, or to an anomalous regime.

iii) the deformation of the anti-BRST symmetry by a non-trivial BRST-exact term will preserve the bicomplex character of the ``dressed'' \cite{12} $\left(M,s_1,s'_2\right)$ bicomplex, only if the system is completely integrable, according to (\ref{7}).

\section{Integrable hierarchies}
Once the integrability and bi- structure are given an interpretation in the BRST-anti-BRST 
framework, one can derive the complete description of such a system, the recursion operator 
and the associated hierarchy.
The celebrated involution property is recovered:
\begin{equation}
\{ H, H_1 \}_{\omega}=\{ H, H_1 \}_{\omega_1}=0
\label{12}
\end{equation}
for the BRST observables associated to the first deformed bicomplex, due to:
\begin{equation}
\{H,H_1\}_{Q,\overline{Q}_1}=\{H,H_1\}_{Q,\overline{Q}}=0
\label{13}
\end{equation}
with the Q-brackets defined in \cite{7}.

The same equation (\ref{7}) gives the recursion operator defined by $T={\omega}^{-1}{\omega}_1$, with vanishing torsion when considered as a tensor field, which yields :
\begin{equation}
\partial_a H_1 = T^{ab} \partial_b H
\label{14}
\end{equation}
or:
\begin{equation}
J\delta H_{n+1}=J_1 \delta H_n
\label{15}
\end{equation}
such that an infinite set of ``deformed'' models can be associated to the original one by the following procedure:
\begin{equation}
H^n_{eff}=\{ Q, \{ \overline{Q} , H_n \} \}=
\{Q, \psi^n\}
\label{16}
\end{equation}
where:
\begin{equation}
\psi^n=\{ \overline{Q} , H_n \}
\label{17}
\end{equation}
due to:
\begin{equation}
\psi^n ={\cal{P}}_a \omega^{ab} \left(T^n\right)^{bc} \partial_c H=
{\cal{P}}_a \omega^{ab} \partial_c H_n
\label{18}
\end{equation}
with:
\begin{equation}
\overline{Q}=-{\cal{P}}_a \lambda_b {\omega}^{ba}-\frac{1}{2}{\cal{P}}_a 
{\cal{P}}_b C^c \partial_c {\omega}^{ba}\left(-\right)^{\epsilon_b \epsilon_c}
\label{19}
\end{equation}

However, the intriguing result recently proved for the KdV hierarchy \cite{14} and concerning 
 the infinite number of symplectic structures finds here an 
interpretation and an algorithmic construction, based on the hierarchy of bicomplexes:
\begin{equation}
\psi^n=\{ \overline{Q}_n, H \}
\label{20}
\end{equation}
with:
\begin{equation}
\overline{Q}_n=\overline{Q}T^n
\label{21}
\end{equation}
and:
\begin{equation}
\{ \phi^a, \phi^b \}_{Q,\overline{Q}_n}=\omega_n^{ab}
\label{22}
\end{equation}

The classical Poincare invariants corresponding to the solutions of $Q\rho=0$ and  
${\overline{Q}}\rho=0$ (if one uses the operator formulation \cite{8}. \cite{9}) are indeed invariant under the Hamiltonian flow $H_{eff}K^p=0$, where $K^p$ are the fundamental BRST-invariant observables.

In conclusion, the interpretation of the integrable hierarchies in terms of BRST-anti-BRST 
formalism is able to provide a criterion for the existence of the bi-Hamiltonian structure 
and to describe the set of models as deformations of a given initial theory, defined by 
$Q,\overline{Q}$ and the recursion operator that connects the tower of 
gauge-fixing ``fermions''.

We will exemplify the proposed formalism, starting from a very well known ``gauge-fixing 
functional'' for $n=1$ ($\phi^1, \phi^2$):
\begin{equation}
H={{\phi}^1}^2+{{\phi}^2}^2
\label{21}
\end{equation}
and a closed two-form $\Omega=d\phi^a\wedge d\phi^b \omega_{ab}$ with:
\begin{equation}
\omega^{12}=-\omega^{21}=-\frac{1}{\hbar}S
\label{22}
\end{equation}
where $S$ is the Schroedinger operator and one identifies $p=\phi^1$ and $q=\phi^2$ as the Darboux coordinates $p\left(x\right)=Im\varphi\left(x\right)$, $q\left(x\right)=Re\varphi\left(x\right)$.

Then the recursion operator exists, $\left(T^{-1}\right)^{11}=\left(T^{-1}\right)^{22}=S$, and for the infinite set of non- anomalous deformations we will derive the corresponding 
integrable hierarchy:

i) for n=1
\begin{equation}
\psi^1={\cal{P}}_a \omega^{ab}_1 \partial_c H
\label{23}
\end{equation}
will generate the equations of motion $\dot{\phi}^a=\{\phi^a,H^1_{eff} \}$ for a=1,2, which is the
first equation in the so-called Schroedinger hierarchy:
\begin{equation}
\dot{\varphi}=-iS\varphi
\label{24}
\end{equation}

ii) for n=2, the next level is recovered, giving the equation:
\begin{equation}
\dot{\varphi}=-iS^2\varphi
\label{25}
\end{equation}
 The higher levels can be constructed then, recursively.
On the opposite, if one chooses to work with the operator $\left(T_{nonlin}^{-1}\right)$ \cite{10} , the nonlinear Schroedinger hierarchy is generated as follows:
\begin{equation}
\dot{\varphi}=\partial_x \varphi
\label{26}
\end{equation}
for n=1, while:
\begin{equation}
\dot{\varphi}=i\left(\partial_{xx}\varphi+|\varphi|^2\varphi\right)
\label{27}
\end{equation}
at n=2, and the iterations may continue.
On the opposite, if one chooses the operator $T_K=\partial_{xx}+\frac{2}{3}\phi+
\frac{1}{3}\phi_xD^{-1}$ (\cite{10}), the KdV hierarchy is obtained.

Consequently, the method based on the BRST-anti-BRST formulation does not provide more then the integrability criterion and the systematic generation of higher levels once the recursion operator is known. However, our analysis may be refined, based on purely cohomological arguments, 
such that to provide the detailed calculation of the second Poisson bracket involved in the 
deformed anti-BRST charge.

\section{Conclusions}
The integrability conditions were derived in this paper for classical mechanics and models with effective Hamiltonians of the type $s_1s_2H$. The bi-Hamiltonian structure was found as a consequence of the existence of a deformation of the gauge-fixing functional and of the anti-BRST charge that leave the effective Hamiltonian unchanged. This in turn allowed us to derive an infinite set of models associated to the chain of ``anomalous'' deformations and the corresponding conserved functionals.   

The present method can be further extended to a ``multi-level'' Hamiltonian formalism, which
 would require new conjugate variables associated to $\chi^a=\omega^{ab}\partial_b H$, and 
higher-level gauge-fixing terms:
\begin{equation}
\psi^{multil}=\sum {\cal{P}}_a^{(n)}\chi^{a(n)}
\label{28}
\end{equation}
providing a different approach to the 
integrability conditions. The treatment of constrained classical systems and specially of the 
ones which are characterised by second-class constraints will be much more complex, but related to this approach, by exploiting the ideas developed in \cite{15} .

A more detailed cohomological study, involving the general structure of a non-trivial 
BRST-invariant term in (\ref{2}), could lead us to the construction of the anti-BRST charge 
deformation and provide a method to calculate $\omega_1$ and the recursion operator.

The main advantage of the BRST formalism is that it may offer an unambiguous and systematic 
quantization method for the bi-Hamiltonian systems, starting with the construction of a 
non-degenerate anti-bracket associated to both Poisson brackets, in an extended version of the 
bracket-antibracket correspondence \cite{16} .

\end{document}